\documentclass[a4paper,12pt]{article}

\usepackage{psfig}
\usepackage{amsmath}
\usepackage{amscd}
\usepackage{amssymb}

\newcommand{\gsim}{\raisebox{-0.5ex}{$\stackrel{>}{\sim}$}}
\newcommand{\lsim}{\raisebox{-0.5ex}{$\stackrel{<}{\sim}$}}

\def\bc{\begin{center}}
\def\ec{\end{center}}

\def\be{\begin{equation}}
\def\ee{\end{equation}}
\def\beq{\begin{eqnarray}}
\def\eeq{\end{eqnarray}}
\def\bfig{\begin{figure}}
\def\efig{\end{figure}}
\def\bnum{\begin{enumerate}}
\def\enum{\end{enumerate}}
\def\noi{\noindent}

\begin{document}
\begin{flushright}
Journal-Ref: Astronomy Letters, 2006, v. 32, No 10, pp. 649-660
\end{flushright}

\vspace{1cm}
\bc
{\baselineskip=24pt
\LARGE {\bf Estimating the Dark Halo Mass from the Relative Thickness of 
Stellar Disks}} \\
\vspace{0.7cm}
{\bf N.Ya.~Sotnikova (nsot@astro.spbu.ru)\\ 
S.A.~Rodionov (seger@astro.spbu.ru)}\\
\vspace{0.7cm}
{\it Sobolev Astronomical Institute, St. Petersburg State University,
St. Petersburg}\\
\ec

\vspace{1.0cm}

We analyze the relationship between the mass of a spherical component and
the minimum possible thickness of stable stellar disks. This relationship
for real galaxies allows the lower limit on the dark halo mass to be
estimated (the thinner the stable stellar disk is, the more massive the dark
halo must be). In our analysis, we use both theoretical relations and
numerical N-body simulations of the dynamical evolution of thin disks in the
presence of spherical components with different density profiles and
different masses. We conclude that the theoretical relationship between the
thickness of disk galaxies and the mass of their spherical components is a
lower envelope for the model data points. We recommend using this
theoretical relationship to estimate the lower limit for the dark halo mass
in galaxies. The estimate obtained turns out to be weak. Even for the
thinnest galaxies, the dark halo mass within four exponential disk scale
lengths must be more than one stellar disk mass. 

\newpage

\section{Introduction}
A large body of observational data provides evidence for the existence of
dark matter, which manifests itself only through its gravitational influence
on stellar systems of various scales (from individual galaxies to clusters
of galaxies and superclusters), determining to a great extent their dynamics
and structure. The controversy surrounding the following two main questions
related to the dark matter is still continuing: What is its nature and to
what extent does it exceed in mass the luminous matter? The first question
is far from being resolved. As regards the second question, on the scales of
individual galaxies, it is formulated as follows: Beginning from which
regions (inner or outer) does the dark halo mass prevail over the luminous
mass?

There are several observational constraints on the mass and extent of the
dark halos in galaxies; these constraints do not depend on what the nature
of the dark matter is. In general, they give an estimate of the lower limit
for the dark halo mass within a fixed radius. For spiral galaxies, this
estimate primarily follows from the analysis of the contributions from
various components of the system to its rotation curve for the so-called
maximum disk model (a classical example of the separation of the
contributions from the disk and the dark halo to the rotation curve of a
galaxy, NGC 3198; van Albada et al. 1985). In this model, the dark halo mass
within the optical radius of a galaxy generally does not exceed the stellar
disk mass. When specific systems are investigated, a joint interpretation of
the data on the rotation curves and radial velocity dispersion profiles,
along with considerations regarding the marginal stability of stellar
disks (see, e.g., Bottema and Gerritsen 1997; Khoperskov et al. 2001; Zasov
et al. 2001; Khoperskov and Tyurina 2003)\footnote{In these papers, the
results of numerical $N$-body simulations with a variable disk mass were
used to explain the observed stellar velocity dispersion for a number of
spiral galaxies.}, often raises significantly the lower limit for the dark
halo mass within a fixed radius (occasionally, interpretation of the data on
the stellar velocity dispersion based on numerical simulations increases the
dark halo mass by almost an order of magnitude compared to its value given
by the maximum disk model\footnote{The galaxy NGC 891 may be cited as an
example (Khoperskov et al. 2001).}).

The upper limit for the dark halo mass is much more difficult to constrain.
Where the region within the optical radius of a galaxy is involved,
stability-related considerations are invoked. For example, the presence of a
regular spiral pattern related to the propagation of density waves in
galaxies requires a gravitationally ``active'' stellar disk, which reduces
the contribution from the spherical component to the total gravitational
field of the spiral galaxy and to the total mass (Athanassoula et al. 1987).
Another important test for the presence of a massive halo is the length of
the tidal tails in interacting spiral galaxies. In interacting systems, we
quite often observe extended tidal features stretching to distances as large
as 40--100 kpc. Based on numerical simulations of interacting disk galaxies
with dark halos that correspond to models obtained in cosmological
calculations (Navarro et al. 1999), Dubinski et al. (1999) showed that the
calculated parameters could be reconciled with the parameters of observed
systems only for two types of models: models with an extended moderate-mass
halo and a rotation curve that is determined almost entirely by the stellar
disk within the optical radius and models with a compact low-mass halo that
makes the main contribution to the rotation curve in the inner disk.
Reconciling the rotation curves of specific interacting systems (e.g., NGC
4038/39 and NGC 7252) and the sizes of their tidal features with the model
parameters leads one to conclude that the stellar disks in such systems
dominate in mass within the optical radius, which gives a strong upper limit
on the dark mass. A similar conclusion was also drawn from the simulations
of the famous Mice galaxy, NGC 4676 (Sotnikova and Reshetnikov 1998).

Comparison of the various estimates for the dark mass in galaxies shows that
the upper and lower limits do not overlap (see, e.g., McGaugh and Block
1998). This gives reason to revise a number of tests, in particular, the
test related to the thickness of the stellar disks in spiral galaxies
suggested by Zasov et al. (1991, 2002). These authors argue that the
thickness of a stable stellar disk at a given halo mass is limited below. If
the disk thickness is smaller than a certain value, then bending instability
will lead to increasing the vertical velocity dispersion and to a thickening
of the system. The growth of bending perturbations is stabilized by a
massive dark halo: the more massive the halo is, the thinner the stable disk
can be. As we showed previously (Sotnikova and Rodionov 2005), not only a
massive dark halo has a stabilizing effect on the growth of bending
perturbations, leaving the stellar disk fairly thin. A low-mass compact
bulge also produces such an effect. In this case, the lower limits on the
dark mass can be mild and consistent with the upper limits. In this paper,
we analyze the relationship between the stellar disk thickness and the dark
halo mass that is derived both from theoretical considerations and from
numerical simulations of the dynamical evolution of thin disks in the
presence of spherical components with different density profiles and
different masses. The derived relationship predicts moderate dark halo
masses even for the thinnest galaxies. 

\section{The stellar disk thickness and the relative mass of the spherical
component}
When the global parameters of stellar disks are analyzed, the disks are
generally assumed to be stable against the growth of perturbations in the
plane and in the vertical direction. The stability conditions for a given
dark halo mass limit below the radial and vertical stellar velocity
dispersions. Since the vertical velocity dispersion is related to the
stellar disk thickness, it is also limited below at a fixed dark halo mass.

The disks are commonly assumed to be at the stability boundary. In this
case, the relative stellar disk thickness and the relative dark halo mass
are related to one another. This relationship was (probably first) noticed
by Zasov et al. (1991) on the following grounds.

For a gravitating layer with a volume density profile that corresponds to an
isothermal (along the $z$ axis) model (Spitzer 1942),
\be
\rho(z)=\displaystyle \frac{\Sigma}{2 z_0} \, {\rm sech}^2(z / z_0) \, ,
\label{eq_dens_sech2}
\ee
the following relation, which is a vertical equilibrium condition, is known
to be valid:
\be
\sigma_z^2 = \pi G z_0 \Sigma \, . 
\label{eq_isothermal_sheet}
\ee
Here, $\Sigma$ is the star surface density of the layer, $z_0$ is the scale
length of the density variations in the $z$ direction, and $\sigma_z^2$ is
the vertical stellar velocity dispersion of the layer. If 
condition~(\ref{eq_isothermal_sheet}) is applied to a stellar disk, then the
$R$ dependences of $\Sigma$, $z_0$, and $\sigma_z$ should be taken into
account. For real stellar disks embedded in a dark halo, the vertical
equilibrium condition must be defined slightly differently (Bahcall 1984).
However, as was shown, for example, by Rodionov and Sotnikova (2006), the
approximation of isothermal layers holds good for the entire disk except the
innermost regions even in the presence of a massive spherical component. 

If the disk is stable against perturbations in the plane, then the
root-mean-square (rms) radial velocity must be
\be
\sigma_R(R) = Q_{\rm T} \, \frac{3.36 \, G \, \Sigma(R)}{\kappa(R)} \, ,
\label{eq_sigma_R_Toomre}
\ee
where $Q_{\rm T} \gsim 1$ is the Toomre parameter,  
$\displaystyle \kappa = \sqrt{2 \left(\frac{v_{\rm c}^2}{R^2} + 
\frac{v_{\rm c}}{R} \frac{d v_{\rm c}}{dR} \right)}$ is the epicyclic
frequency, and $v_{\rm c}$ is the linear circular velocity. The coefficient
3.36 in Eq.(\ref{eq_sigma_R_Toomre}) corresponds to an infinitely thin model
disk (Toomre 1964).

An expression for the minimum possible relative thickness of a stable disk
can be derived from equilibrium condition (\ref{eq_isothermal_sheet}) and
stability condition (\ref{eq_sigma_R_Toomre}). To be able to compare
theoretical estimates with observational data, let us write the expression
for the relative disk thickness at a certain radius, for example, $R = 2h$,
where $h$ is the exponential radial disk scale length, which enters into the
commonly used law of surface density variations with radius:
$$
\Sigma(R) = \displaystyle \frac{M_{\rm d}}{2 \pi h^2} \exp{(-R/h)} \, ,
$$ 
where $M_{\rm d}$ is the total disk mass. The radius $2h$ was chosen on the
grounds that the observed vertical disk structure in this region is
generally no longer distorted by the central bulge, the bar, and the
so-called X-shaped structures. 

At $R \gsim 2h$, the rotation curves of luminous galaxies generally reach a
plateau. In this case, the epicyclic frequency (see stability
condition~(\ref{eq_sigma_R_Toomre})) is related to the linear circular
velocity $v_{\rm c}$ by $\kappa(R) = \sqrt{2} v_{\rm c} / R$. In the plateau
region, $v_{\rm c}$ is roughly constant and, hence, $v_{\rm c}(2h) \approx
v_{\rm c}(4h)$. We can then use $v_{\rm c}$ to estimate the total mass of a
galaxy (including the mass of its spherical component $M_{\rm sph}$) within
a fixed radius, for example, $R=4h$, $M_{\rm tot}(4h) \approx 4 h v_{{\rm
c}}^2 / G$ (see, e.g., Zasov et al. 2002). In addition, the marginal
stability condition in the disk plane allows $Q_{\rm T}$ to be chosen almost
uniquely. Based on numerical $N$-body simulations, Khoperskov et al. (2003)
showed that $Q_{\rm T}$ at $R \simeq (1-2)h$ for marginally stable 3D
stellar disks is equal to a nearly constant value of $1.2 - 1.5$, which does
not depend on the relative mass of the spherical component. Based on the
results of Khoperskov et al. (2003), we took $Q_{\rm T}(2h) \approx 1.4$.
Using Eqs.(\ref{eq_isothermal_sheet})~and~(\ref{eq_sigma_R_Toomre}), under
our additional assumptions, we then obtain
\be
\frac{z_0(2h)}{h}
\approx 1.2 \, \frac{\sigma_z^2}{\sigma_R^2} \, 
\frac{M_{\rm d}}{M_{\rm tot}(4h)} \, .
\label{eq_z-M}
\ee

If we fix the ratio $\sigma_z / \sigma_R$ at the level given by the local
linear criterion for bending instability, i.e., at approximately $0.29 -
0.37$ (Toomre 1966; Kulsrud et al. 1971; Polyachenko and Shukhman 1977;
Araki 1985), then, according to Eq.(\ref{eq_z-M}), the relative stellar disk
thickness measured at the given distance $R = 2h$ is 
\be
\frac{z_0(2h)}{h} \approx
\frac{0.18}{1 + \mu(4h)} \, ,
\label{eq_z-M_local}
\ee 
where $\mu(4h) = M_{\rm sph}(4h) / M_{\rm d}(4h)$, $M_{\rm tot}(4h) = M_{\rm
d}(4h) + M_{\rm sph}(4h)$, $M_{\rm sph}(4h)$ -- mass of spheroidal component.
In~(\ref{eq_z-M_local}), we took into account the fact that $90\%$ of the
disk mass is contained within $4h$ for an exponential surface density
profile. The numerical coefficient in Eq.(\ref{eq_z-M_local}) was obtained
for $\sigma_z / \sigma_R = 0.37$ (this value corresponds to the linear
criterion for bending instability in Polyachenko and Shukhman (1977)). The
larger the relative mass of the spherical component within a fixed radius,
for example, $4h$, the smaller the relative disk thickness $z_0(2h) / h$ at
this radius. The latter conclusion is one of the main conclusions reached by
Zasov et al. (1991). In the conclusion under discussion the coefficient in
relation~(\ref{eq_z-M_local}) will be of interest in the subsequent analysis.

Equation~(\ref{eq_z-M_local}) was derived for the specific law of density
variations in the vertical direction (\ref{eq_dens_sech2}). In certain
cases, the vertical volume density profile of the disk matter is fitted by
other laws (van der Kruit 1988), for example, 
\be
\rho(z)=\displaystyle \frac{\Sigma}{\pi z_e} \, {\rm sech}(z / z_e)
\label{eq_dens_sech1}
\ee
or an exponential law,
\be
\rho(z)=\displaystyle \frac{\Sigma}{2 h_z} \, \exp(- |z| / h_z) \, ,
\label{eq_dens_exp}
\ee
where $z_e$ and $h_z$ are the corresponding vertical scale heights that can
depend on $R$. It can be shown that an equilibrium condition
similar~(\ref{eq_isothermal_sheet}) for profiles~(\ref{eq_dens_sech1})
or~(\ref{eq_dens_exp}) may be written as
\be
<\sigma_z^2> = C \pi G \Delta \Sigma \, ,
\label{eq_isothermal_disk}
\ee
where $<\sigma_z^2>$ where is the vertical velocity dispersion averaged
along the $z$ axis; $\Delta$ is the half-thickness of a homogeneous layer
with a density equal to the density in the disk at $z = 0$; and $C = 1$, $C
= 28 \, \zeta(3) / \pi^3 \approx 1.0854$ ($\zeta$ is the Riemann function),
and $C = 1.5$, respectively\footnote{To obtain these coefficients, we must
substitute the corresponding laws of density variations in the Jeans
equation that describes the vertical equilibrium and, having calculated the
law of velocity dispersion variations along $z$, average this quantity.},
for density profiles~(\ref{eq_dens_sech2}),~(\ref{eq_dens_sech1}),
and~(\ref{eq_dens_exp}). Below, we argue that the vertical profiles of model
disks at late evolutionary stages are more likely described by
laws~(\ref{eq_isothermal_sheet})~or~(\ref{eq_dens_sech1}) than
by~(\ref{eq_dens_exp}). In this case, $C \approx 1.0-1.1$ in equilibrium
condition~(\ref{eq_isothermal_disk}) and the numerical coefficient
in~(\ref{eq_z-M_local}) is virtually constant if $\Delta$, the
half-thickness of a homogeneous layer\footnote{For density
profile~(\ref{eq_isothermal_disk}), the vertical scale height $z_0$ is
exactly equal to the half-thickness of a homogeneous layer $\Delta$.}, is
used in place of $z_0$.

Note that relation~(\ref{eq_z-M_local}) is valid for a marginally stable
stellar disk. If a galaxy has a margin of stability against perturbations in
the plane or against bending perturbations, then the greater sign must be
substituted for the sign of approximate equality in~(\ref{eq_z-M_local}). As
a result, for stable stellar disks, Eq.(\ref{eq_z-M_local}) yields
\be
\mu(4h) \, \gsim \, 0.18 \frac{h}{z_0(2h)} - 1 \, .
\label{eq_M-z}
\ee

The lower limit for the mass of the spherical component in a galaxy, in
particular, the dark halo mass, can be estimated from the observed relative
thickness of the galaxy using relation~(\ref{eq_M-z}). Note that the
theoretical estimate is weak. Even for extremely thin galaxies with $z_0 / h
= 1/15$, we find that $\mu(4h) \gsim 1.7$. For example, for the thinnest
galaxy in the sample that was used by Zasov et al. (2002), $z_0/h \approx
1/11$ (see Fig.4b in this paper). For such a galaxy, according
to~(\ref{eq_M-z}), $\mu(4h) \gsim 0.98$. In the sample of edge-on galaxies
whose parameters were discussed by Kregel et al. (2002), $z_0/h \approx
1/9.7$ for the thinnest galaxy (435-G25) (see Table 1 in this
paper)\footnote{In this paper, exponential profile~(\ref{eq_dens_exp}) was
used to fit the vertical stellar disk density profile. The vertical scale
height was defined by $h_z$, which, according to the authors themselves, is
related to $z_0$ for density profile~(\ref{eq_dens_sech2}) by $h_z = 0.5
z_0$.}. For such a galaxy, Eq.(\ref{eq_M-z}) yields $\mu(4h) \gsim 0.75$.

\section{Numerical simulations}
The relationship between the relative stellar disk thickness and the
relative dark halo mass derived from five N-body simulations was also first
discussed by Zasov et al. (1991). A similar relationship was constructed by
Mikhailova et al. (2001) using several tens of simulations. In general, the
two relationships coincide.

It should be noted that theoretical relation~(\ref{eq_z-M_local}) written
above was derived from the considerations that were used to substantiate the
results of $N$-body simulations (Zasov et al. 1991, 2002). However, even a
superficial comparative analysis shows that the curve corresponding to
theoretical relation~(\ref{eq_z-M_local}) lies well below the curve
constructed from the experimental data points (see Fig.2 in Mikhailova et
al. (2001)). According to the analytical studies of various authors, the
factor $\sigma_z / \sigma_R$ appearing in Eq.(\ref{eq_z-M}) lies within the
range $0.29 - 0.37$. We took the maximum value of this factor, 0.37. If we
took a smaller velocity dispersion ratio, then the differences between the
theoretical curve and the experimental data points would be even larger. 
 
The relationship analyzed by Zasov et al. (1991) and Mikhailova et al.
(2001) can be roughly approximated by the formula $(0.3 - 0.35)/(1 +
\mu(4h))$. Thus, the thickness of the numerical stellar disk models is
approximately twice the thickness calculated using~(\ref{eq_z-M_local}). In
other words, the relationship derived from $N$-body simulations predicts a
higher dark halo mass than estimate~(\ref{eq_M-z}). In our view, this
discrepancy makes it difficult to interpret both the results of numerical
simulations and the relationships derived from observational data (Zasov et
al. 2002) and requires a more refined analysis of the numerical simulations
that primarily includes the stabilizing role of a compact low-mass spherical
component (Sotnikova and Rodionov 2005).   

\subsection{Numerical Models}
\label{s_models}
Below, we analyze the parameters of several tens of models that we obtained
from numerical $N$-body simulations. We used an algorithm based on the
hierarchical tree algorithm (Barnes and Hut 1986) and some of the codes from
the NEMO software package (http:\/\/astro.udm.edu/nemo; Teuben 1995). 

The stellar disk was represented by a system of $N$ gravitating bodies ($N$
was varied within the range $300\, 000$ to $500\, 000$) and the spherically
symmetric component was described in terms of an external static potential
that was a superposition of two potentials, the bulge and the dark halo.   

The initial particle velocities (the rotational and random velocities) were
specified on the basis of equilibrium Jeans equations using a standard
technique (see, e.g., Hernquist 1993). A description of the method for
constructing the initial model can be found in our previous papers
(Sotnikova and Rodionov 2003, 2005). Also given there are the details
concerning the bulge model (a Plummer sphere) and the halo model (a
logarithmic potential).

The ratio of the mass of the spherical component to the disk mass within
$4h$ was varied between 0.25 and 4. The mass of the spherical component was
distributed differently between the bulge and the halo. The scale length of
the bulge was chosen to be $a_{\rm b} = h / 7$ almost for all models. We
also considered models with concentrated ($a_{\rm h} = h / 1.75$) and
``loose'' ($a_{\rm h} = h / 0.35$) halos, $a_{\rm h}$ is the scale length of
the mass distribution in the halo. The potential softening length $\epsilon$
for all models corresponded to our recommendations given previously
(Rodionov and Sotnikova 2005), $\epsilon = h / 175$. 
  
For all models, we chose nearly equilibrium (in the radial and vertical
directions), but unstable (against the growth of bending perturbations)
initial conditions. During the growth of bending instability, the system
was dynamically heated, the velocity dispersion increased in the $z$
direction, and the disk thickness increased to a value close to the lower
boundary of a steady state. The main parameter in which the stellar disk
models being analyzed differ is the fraction of the total mass of the galaxy
contained in its spherical component (the dark halo and the bulge).

\subsection{Determining the stellar disk thickness in numerical simulations}
\label{s_thickness}
Before discussing in detail the relationship between the relative stellar
disk thickness and the relative dark halo mass in galaxies, we should gain
an understanding of how the thickness of numerically simulated disks is
determined. For model disks, the thickness at a given distance $R$ from the
disk center is usually estimated as the rms deviation of the $z$ coordinates
of stars from the symmetry plane of the disk (see, e.g., Khoperskov and
Tyurina 2003):
\be
z_{\rm rms} = \sqrt{\displaystyle 
\left<\left(z-<z>\right)^2\right>} \, .
\label{eq_z_rms}
\ee

The disk in the $z$ direction is commonly assumed to have volume density
profile~(\ref{eq_dens_sech2}) that corresponds to an isothermal model. In
this case, $z_0$ and $z_{\rm rms}$ are related by
\be
z_0 = \frac{2 \sqrt{3}}{\pi} \, z_{\rm rms} \approx 1.10 \, z_{\rm rms} \, .
\label{eq_z0_rms}
\ee

For density profile~(\ref{eq_dens_sech2}), $z_0$ is the half-thickness of a
homogeneous layer, $\Delta$, with the volume density at $z = 0$. The
parameter $z_0$ is estimated from observational data for real galaxies by
analyzing how the surface brightness of the edge-on stellar disk varies
along the z axis and by fitting the brightness distribution to $\propto {\rm
sech}^2(z/z_0)$. Based on~(\ref{eq_z0_rms}), one can assume that the
parameter $z_{\rm rms}$ determined from~(\ref{eq_z_rms}) for numerical
models quantitatively differs only slightly from the disk half-thickness
$z_0$ appearing in law~(\ref{eq_dens_sech2}) and characterizes the thickness
of real disk galaxies. However, we will now show that in practice $z_{\rm
rms}$ is a poor characteristic of the model disk thickness and that using it
can lead to incorrect relations between the stellar disk thickness and other
global parameters of spiral galaxies.
 
A certain number of particles that fly far away from the disk plane appear
in a model galaxy during its evolution. Since their number is small, the $z$
particle distribution is generally well fitted by isothermal density
profile~(\ref{eq_dens_sech2}). However, this small number of distant
particles contributes significantly to $z_{\rm rms}$ and leads to an
overestimation of the stellar disk thickness calculated in this way. The
extent to which the thickness is overestimated depends on the relative mass
of the spherical component and on the distance from the center at which we
measure the thickness.

Let us introduce another parameter that can characterize the thickness of
model stellar disks at a given distance $R$: $z_{1/2}$ --- the median of the
absolute value of $z$. Twice $z_{1/2}$ is the thickness of the layer within
which one half of the particles is contained. For density
profile~(\ref{eq_dens_sech2}), based on the introduced definitions, we have
\beq
\Delta &=& z_0 \, , \nonumber \\ 
\frac{\Delta}{2 z_{1/2}} &=& \frac{1}{\ln 3} \approx 0.91  \, ,
\label{eq_ratio_sech2}\\
\frac{z_{\rm rms}}{2 z_{1/2}} &=& \frac{\pi}{2 \sqrt{3} \ln 3} 
\approx 0.83 \, . \nonumber
\eeq

It is easy to show that the relations between $z_{1/2}$, $\Delta$, and
$z_{\rm rms}$ for density profile~(\ref{eq_dens_sech1})
and~(\ref{eq_dens_exp}) can be written as follows: for~(\ref{eq_dens_sech1}),
\beq
\Delta &=& z_{\rm rms} = \frac{\pi}{2} \, z_e \, , \nonumber \\
\frac{\Delta}{2 z_{1/2}} &=& 
\frac{\pi}{4 \ln\left(\tan\displaystyle\frac{3 \pi}{8}\right)} 
\approx 0.89 \, , 
\label{eq_ratio_sech1} \\
\frac{z_{\rm rms}}{2 z_{1/2}} & \approx& 0.89 \, , \nonumber
\eeq
and for~(\ref{eq_dens_exp}),
\beq
\Delta &=& \frac{\sqrt{2}}{2} \, z_{\rm rms} = h_z \, , \nonumber \\ 
\frac{\Delta}{2 z_{1/2}} &=& \frac{1}{2 \ln 2} \approx 0.72 \, ,
\label{eq_ratio_exp} \\
\frac{z_{\rm rms}}{2 z_{1/2}} &=& \frac{1}{\sqrt{2} \ln 2} \approx 1.02 \, . 
\nonumber
\eeq

\begin{figure}
\centerline{\psfig{file=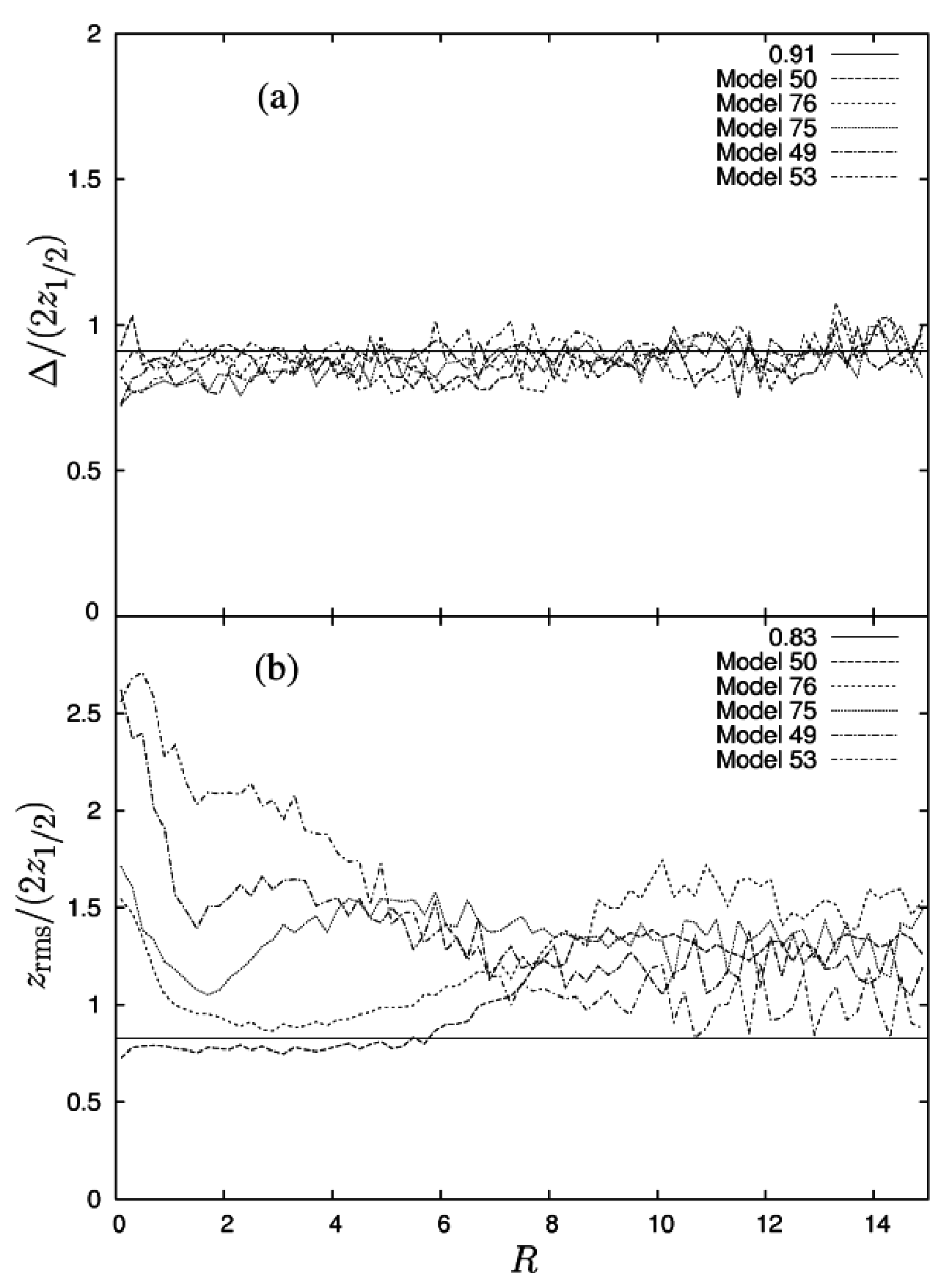,width=12cm}}
\caption[]{
(a) $\Delta / (2 z_{1/2})$ vs. R for several models. The solid line
indicates the 
theoretical value of 0.91 for this ratio for density 
profile~(\ref{eq_dens_sech2}). All models
have the same mass of the spherical component and differ only in mass
distribution of this component between the bulge and the halo (the models
are listed in order of increasing bulge mass fraction). For all models,
$M_{\rm sph}(4h)/M_{\rm d}(4h) = 0.5$ and $h = 3.5$ kpc. 
(b) $z_{\rm rms} / (2 z_{1/2})$ vs. R for the same
models. The solid line indicates the theoretical value of 0.83 for this
ratio for density profile~(\ref{eq_dens_sech2}).}
\label{fig_thickness_halo}
\end{figure}

\begin{figure}
\centerline{\psfig{file=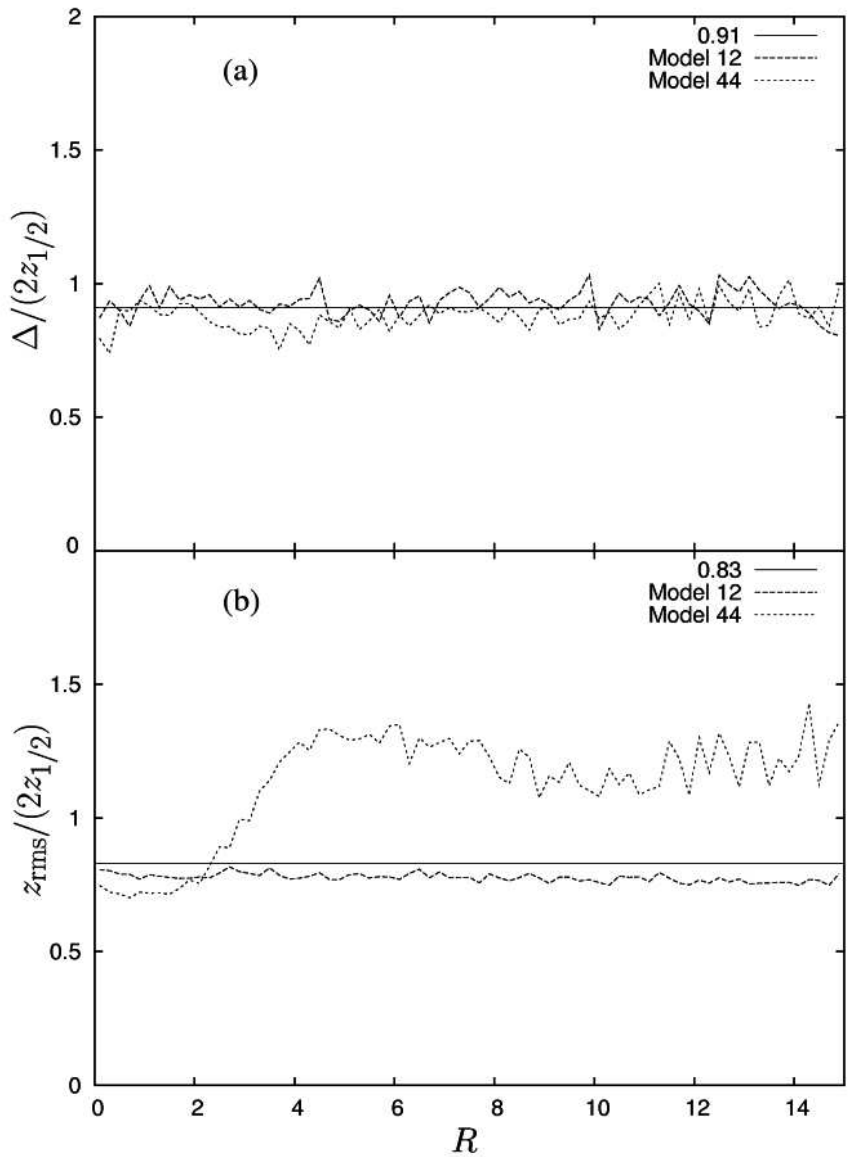,width=12cm}}
\caption[]{
Same as Fig.~\ref{fig_thickness_halo} for two other models -- with relatively
massive (model 44) and very massive (model 12) spherical components. These
models are bulgeless. For
models 44 and 12,  $\mu(4h) = 1.5$ and $\mu(4h) = 3.43$, 
respectively; $h = 3.5$~kpc.}
\label{fig_thickness_bulge}
\end{figure}

These relations allow us to check how adequately the introduced parameters
describe the model disk thickness and to reach certain conclusions regarding
the law by which the vertical density profile can be fitted.
Figures~\ref{fig_thickness_halo}~and~\ref{fig_thickness_bulge} show the
azimuthally averaged radial profiles of the ratios $\Delta / 2 z_{1/2}$ and
$z_{\rm rms} / 2 z_{1/2}$ for several dynamical stellar disk models.

We see from Figs.~\ref{fig_thickness_halo}a~and~\ref{fig_thickness_bulge}a
that the ratio $\Delta / 2 z_{1/2}$ is almost constant along the radius for
various models. Note that the particle parameters from different parts of
the disk are used to determine $z_{1/2}$, $\Delta$, and $z_{\rm rms}$. When
$z_{1/2}$ is estimated, it does not matter how extended the tail of distant
particles is, but the number of particles that contribute to $z_{1/2}$ is
large and accounts for exactly one half of all particles. The volume density
produced by only those particles that are located near the main plane of the
galaxy and the total surface density are used to calculate $\Delta$; the
positions of all the remaining particles do not affect in any way the
determination of $\Delta$. As regards $z_{\rm rms}$, all particles,
including those that randomly flew far away from the disk plane, contribute
to this quantity. It follows from the constancy of the ratio $\Delta / 2
z_{1/2}$ that these two quantities are adequate estimates of the model disk
thickness; otherwise, it would be difficult to explain the coherent behavior
of these quantities. At the same time, the equality of this ratio to 0.9 is
an argument for the fact that the vertical density profile of stellar disks
is similar to law~(\ref{eq_dens_sech2}) or~(\ref{eq_dens_sech1}).      

As follows from
Figs.~\ref{fig_thickness_halo}b~and~\ref{fig_thickness_bulge}b, $z_{\rm rms}
/ 2 z_{1/2}$ varies significantly along the radius $R$ for all of the
analyzed models and, in addition, this ratio is almost always systematically
higher than that given by theoretical
relations~(\ref{eq_ratio_sech2}),~(\ref{eq_ratio_sech1}), and
even~(\ref{eq_ratio_exp}). The thickness overestimation strongly depends on
the degree of matter concentration in the spherical component and on its
relative mass. It is different in different parts of the disk. The
differences are largest for the models with bulges. For them, the disk
thickness in a region of the order of two exponential disk scale lengths can
be overestimated by 100\% or more (models~49~and~53,
Fig.~\ref{fig_thickness_halo}b) if $z_{\rm rms}$ is used. As regards regions
of more than two radial disk scale lengths, the thickness is overestimated
here by 50\% or more. For the models with a relative mass of the spherical
component $\mu(4h) \leq 1.5$ without a bulge (models~50~and~44,
Figs.~\ref{fig_thickness_halo}b~and~\ref{fig_thickness_bulge}b), the
thickness of the central regions is more or less adequately described by the
parameter $z_{\rm rms}$. In this case, the density profile corresponds to
isothermal law~(\ref{eq_dens_sech2}), for which $z_{\rm rms} / 2 z_{1/2}
\approx 0.8$. However, at distances larger than one or one and a half radial
exponential disk scale lengths, the thickness proves to be grossly
overestimated, approximately by 50\%. Only for the models with $\mu(4h)
\gsim 3.0$ (model~12, Fig.~\ref{fig_thickness_bulge}b) does $z_{\rm rms}$
serve as a good estimate of the thickness. In addition, the $z_{\rm rms}$
fluctuations along $R$ for the model disks are too large even when a large
number of particles is used, in contrast to $z_{1/2}$ (Sotnikova and
Rodionov 2005). The ratio $z_{\rm rms} / 2 z_{1/2}$ also exhibits the same
large fluctuations.

\begin{figure}
\centerline{\psfig{file=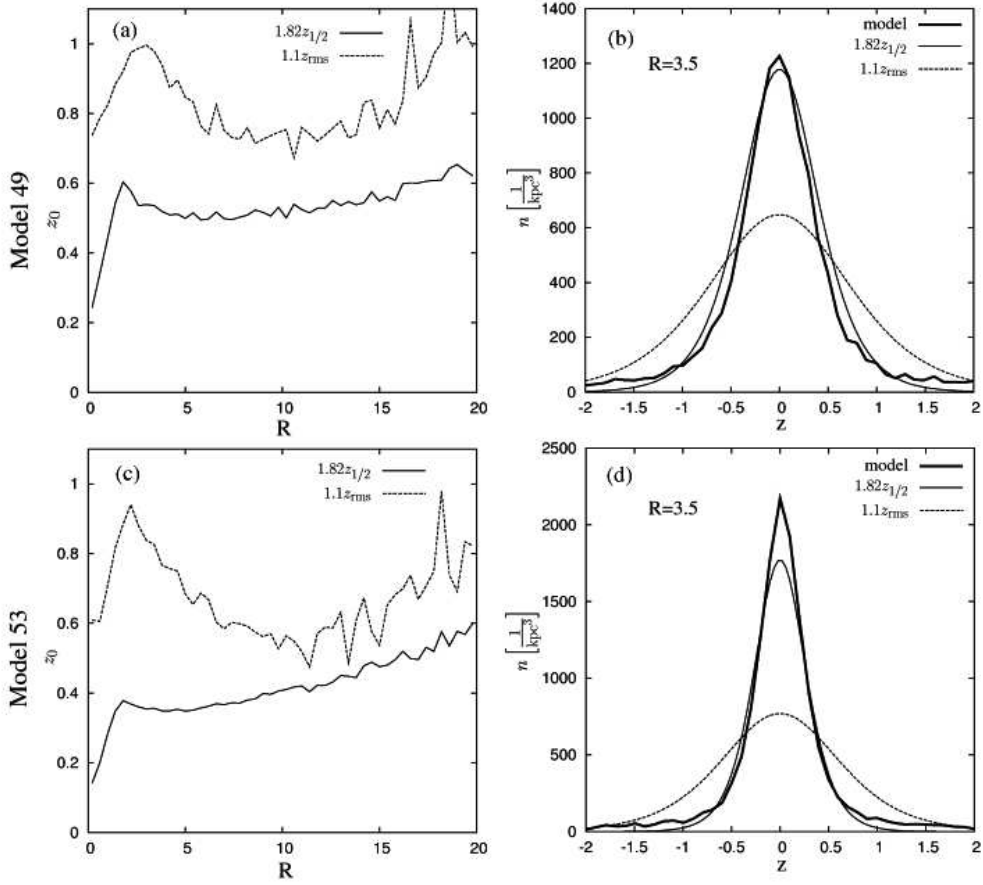,width=14cm}}
\caption[]{
Comparison of the disk thicknesses calculated using $z_{\rm rms}$ and
$z_{1/2}$ for two models at the time $t = 5000$~Myr. Models 49 and 53 have
$M_{\rm sph}(4h) / M_{\rm d}(4h)=0.5$, where $h = 3.5$ kpc, but differ in
bulge mass fraction (it is larger for model 53).
(a), (c) $z_0$~vs.~R; the solid and dashed lines
represent $z_0$ estimated as $1.82 \, z_{1/2}$ and $1.10 \, z_{\rm rms}$,
respectively. (b), (d)
Vertical particle density profiles at the radius $R = 3.5$ kpc; the thick
solid line represents the profile obtained from the model; the thin solid
line represents profile~(\ref{eq_dens_sech2}) in which $z_0$ was estimated
via $z_{1/2}$; the dashed line represents profile~(\ref{eq_dens_sech2}) in
which $z_0$ was estimated via $z_{\rm rms}$. The disk 
surface density  at a given distance $R$ in Eq.~(\ref{eq_dens_sech2})
was calculated from the model.}
\label{fig_z0_49_53}
\end{figure}

The parameter $z_0$ is commonly used to analyze observational data by
assuming the vertical density profile to be isothermal. For our numerical
models, we can determine $z_0$ via both $z_{\rm rms}$ ($z_0 \approx 1.10 \,
z_{\rm rms}$) and $z_{1/2}$ ($z_0 \approx 1.82 \, z_{1/2}$). As we see from
the plots of the radial disk thickness profile
(Figs.~\ref{fig_z0_49_53}a~and~\ref{fig_z0_49_53}c), $z_0$ calculated via
$z_{\rm rms}$ is considerably larger than $z_0$ calculated via $z_{1/2}$.
Figures~\ref{fig_z0_49_53}b~and~\ref{fig_z0_49_53}d show the vertical
density profiles at a given radius ($R = 3.5$ kpc, which corresponds to the
exponential radial disk scale length) and the fit to the density profile by
isothermal law~(\ref{eq_dens_sech2}) with the parameter $z_0$ calculated as
$z_0 \approx 1.10 \, z_{\rm rms}$ and as $z_0 \approx 1.82 \, z_{1/2}$. We
see that if $z_0$ is determined via $z_{1/2}$, then
law~(\ref{eq_dens_sech2}) fits excellently the vertical density profile in the
model. When $z_0$ is determined via $z_{\rm rms}$, the discrepancies between
the profiles are enormous.

The above arguments lead us to the following conclusion: the parameter
$z_{1/2}$ should be taken as the stellar disk thickness to analyze the
relationships between the disk thickness and other global parameters of
galaxies in numerical models. Using $z_{\rm rms}$ can distort significantly
the relationship between the relative thickness of a marginally stable
stellar disk and the relative mass of the spherical component and can raise
the lower limit for the dark halo mass. Below, we use $1.82 \, z_{1/2}$,
since it corresponds to $z_0$ for density profile~(\ref{eq_dens_sech2}).

\subsection{Analysis of numerical simulation results}
In deriving Eq.~(\ref{eq_z-M_local}) for the ratio of the vertical and
radial velocity dispersions, we used the value given by the local linear
criterion (Toomre 1966; Kulsrud et al. 1971; Polyachenko and Shukhman 1977;
Araki 1985). However, the saturation level of the bending instability in
numerical simulations can be considerably higher than this value (see, e.g.,
Sellwood and Merritt 1994; Merritt and Sellwood 1994; Sotnikova and Rodionov
2003). This probably explains why the relationship in Zasov et al.(1991) and
Mikhailova et al.(2001) lies above the ``theoretical'' one. 

The ratio $\sigma_z / \sigma_R$ at which the bending instability is
saturated depends primarily on whether the system admits the growth of
global modes. If the growth of global modes is suppressed, then the ratio
$\sigma_z / \sigma_R$ at a given radius is determined by the local
conditions and is equal to its value that follows from the linear criterion.
If, however, global modes are possible in the system, then the level of
vertical disk heating can be considerably higher.

\begin{figure}
\centerline{\psfig{file=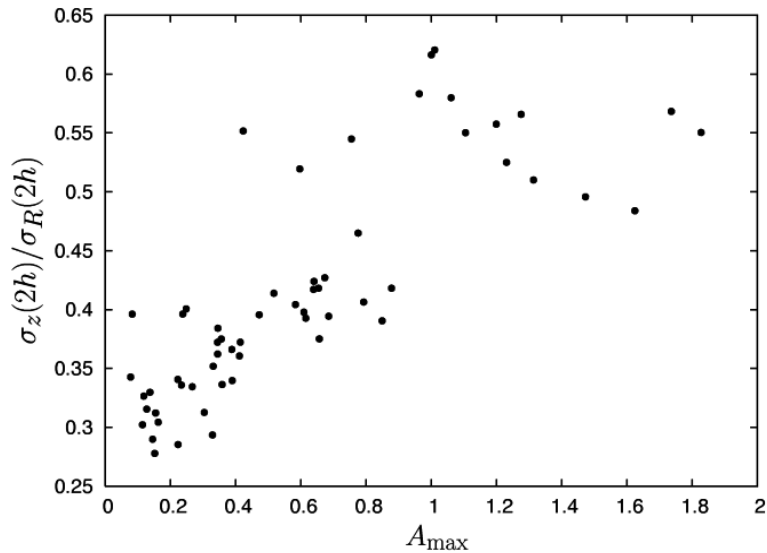,width=11cm}}
\caption[]{
$\sigma_z(2h) / \sigma_R(2h)$ determined at the time 
$t = 3000$~Myr~vs.~maximum disk 
warp amplitude $A_{\rm max}$ for the period from 0 to 3000 Myr (the disk
region within 10 kpc was considered).}
\label{fig_amplituda_max}
\end{figure}

Sweeping throughout the disk, the global modes cause the ratio $\sigma_z /
\sigma_R$ to increase to $0.6$ and, in central disk regions, to $0.7-0.9$
(Sotnikova and Rodionov 2003). In Fig.~\ref{fig_amplituda_max}, the ratio
$\sigma_z / \sigma_R$ is plotted against the maximum amplitude of bending
modes in the time interval from 0~to~3000 Myr for the distance $R = 2h$. The
plot was constructed using numerical simulations of disk galaxies. We see a
clear correlation between the maximum amplitude of the growing and
propagating (through the disk) perturbation and the instability saturation
level and, accordingly, the final $\sigma_z(2h) / \sigma_R(2h)$. When global
modes are possible in the system, the maximum perturbation amplitude turns
out to be larger.

For inhomogeneous disks, the possibility of the growth and propagation of
global modes is determined by the conditions in the central disk regions
(Sellwood 1996; Sotnikova and Rodionov 2005). These conditions impose
constraints on the vertical oscillation frequencies related to the spherical
component and the disk and for the model with $\mu(4h) \lsim 2.0$ depend not
only on $\mu$, but also on the degree of matter concentration in the
spherical component toward the center (Sotnikova and Rodionov 2005). A
low-mass compact bulge suppresses the growth of global bending modes as
effectively as a massive dark halo with a large scale length of the matter
distribution.

\begin{figure}
\centerline{\psfig{file=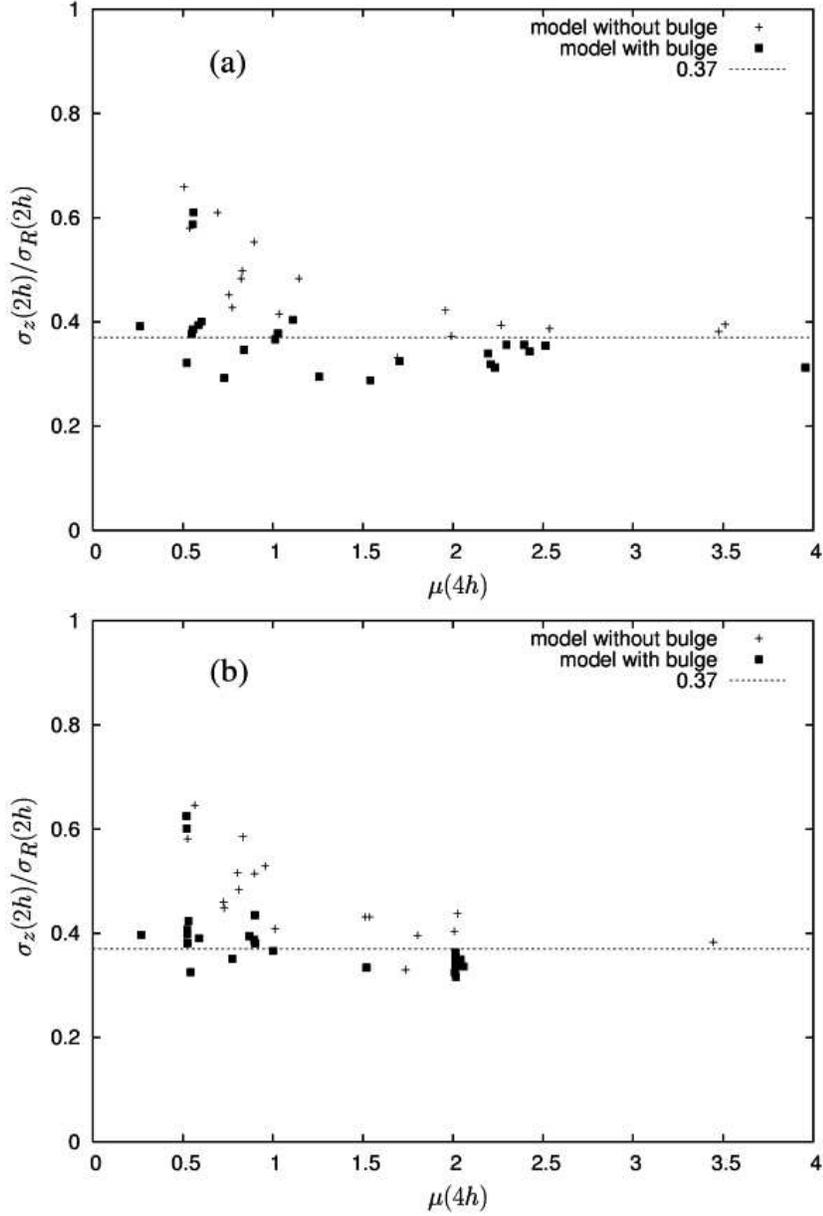,width=11cm}}
\caption[]{ 
$\sigma_z(2h)/\sigma_R(2h)$~vs.~$\mu(4h)$ for our models at the time 
$t = 5000$~Myr. The 
crosses and squares indicate the models without and with a bulge,
respectively. The dashed line indicates $\sigma_z / \sigma_R = 0.37$ that
corresponds to the 
saturation level of the bending instability derived from the linear
criterion of Polyachenko and Shukhman (1977). (a) The exponential disk scale
length $h$ was calculated from the model by the least-squares method; the disk
region of $R \ge 5$ kpc and $R \le 10$ kpc was used to calculate $h$. 
(b) The initial exponential disk scale length $h = 3.5$ kpc was taken as $h$.}
\label{fig_svz_svR_Mh_all}
\end{figure}

In Fig.~\ref{fig_svz_svR_Mh_all}a, $\sigma_z(2h) / \sigma_R(2h)$ is plotted
against the relative mass of the spherical component $\mu(4h)$ for a stellar
disk that came to equilibrium after the saturation of the bending
instability ($t = 5000$~Myr). It should be noted that a difficulty with the
determination of the exponential stellar disk scale length $h$ arises when
constructing this plot. During the evolution, an appreciable mass
redistribution takes place in many models. The density in the central region
can increase greatly. The most significant changes are observed in the models
with the formation of a bar. As a result, the density profile in these
central regions was fitted by an exponential law with a considerably smaller
scale length than that of the density profile on the periphery. For our
analysis, we took the exponential scale length estimated from the periphery
of a model disk (the region from $5$ to $10$ kpc),
Fig.~\ref{fig_svz_svR_Mh_all}a. For comparison,
Fig.~\ref{fig_svz_svR_Mh_all}b shows the same plot, but the value of $h$ was
taken at the initial time ($h = 3.5$~kpc for all models at the initial
time). The total mass within $4h$ also varies with $h$. This explains the
differences in the ranges of $\mu(4h)$ for the same set of models  in
Figs.~\ref{fig_svz_svR_Mh_all}a~and~\ref{fig_svz_svR_Mh_all}b. The squares
in both figures indicate the data for the models with a bulge whose mass is
several times lower than the disk mass. Irrespective of $\mu(4h)$, the
minimum value of $\sigma_z(2h)/\sigma_R(2h)$ obtained from the models
satisfies the linear criterion for bending instability, $(0.29 - 0.37)$. This
minimum value corresponds to the models with a compact bulge or a massive
halo.

The stabilizing role of a compact bulge lies in the fact that it inhibits
the growth of large-scale high amplitude perturbations. In the models with a
low mass spherical component and without a compact bulge, the processes that
may be arbitrarily called a ``violent'' stage of the bending instability
proceed during the evolution (for the appearance of a bell mode in hot
models with large $Q_{\rm T}$ and the rapid growth of bending instability of
the bar in cold disks, see Sotnikova and Rodionov (2003)). The physics of
these processes may not be directly related to the physics of the bending
instability. A vertical resonance can appear in such models (Combes and
Sanders 1981; Combes et al.1990; Pfenniger and Friedli 1991; Patsis et
al.2002). Nevertheless, these processes affect $\sigma_z/\sigma_R$ --- it
becomes considerably higher than that given by the linear criterion for bending
instability.

\begin{figure}
\centerline{\psfig{file=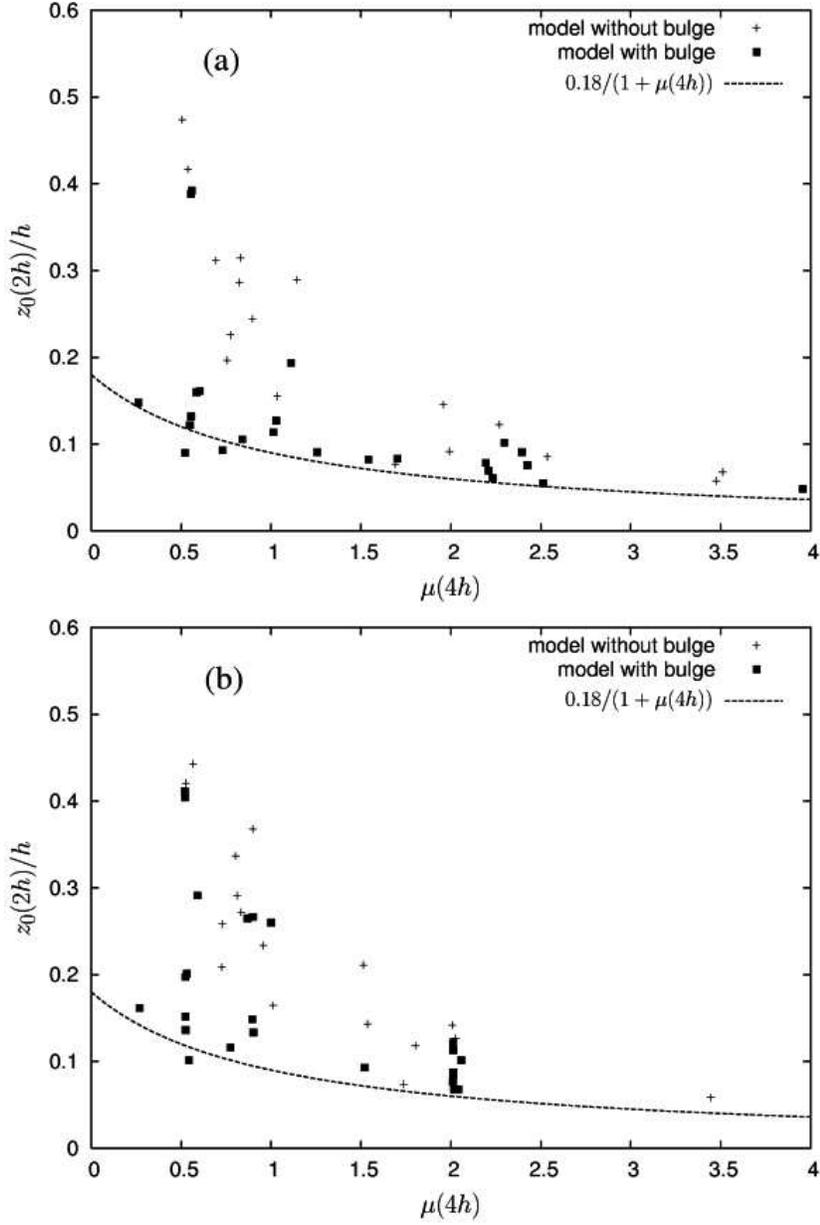,width=11cm}}
\caption[]{ 
Relative stellar disk thickness at radius $2h$~vs.~relative mass of
the spherical component within $4h$ for our models at the time $t = 5000$~Myr.
The value of $z_0$ was calculated as $1.82 \, z_{1/2}$.
The crosses and squares indicate the models without and with a bulge,
respectively. The dashed line represents theoretical relation 
(\ref{eq_z-M_local}). (a) The exponential disk scale length $h$
was determined from the model by the least-squares method; the disk region
of $R \ge 5$ kpc and $R \le 10$ kpc was used to calculate $h$. 
(b) The initial exponential disk scale length $h = 3.5$ kpc was taken as $h$.}
\label{fig_thickness_Mh_all}
\end{figure}

In Fig.~\ref{fig_thickness_Mh_all}a, $z_0(2h)/h$ is plotted against
$\mu(4h)$ at the time $t = 5000$~Myr for $h$ calculated from the region from
$5$ to $10$~kpc. Figure~\ref{fig_thickness_Mh_all}b shows the same plot for
an initial value of $h = 3.5$~kpc. As expected, at fixed relative mass of the
spherical component $\mu(4h)$, the models with bulges can be considerably
thinner than the models without bulges
(see~Figs.~\ref{fig_thickness_Mh_all}a~and~~\ref{fig_thickness_Mh_all}b).
This effect is most pronounced for the models with a spherical component of
low relative mass ($\mu(4h) \lsim 1$).

The following result proved to be more unexpected. The analytical relation
between the minimum possible $z_0(2h)/h$ for a stable disk and $\mu(4h)$
(see~(\ref{eq_z-M_local})) is a lower envelope for the model data
(see~Figs.~\ref{fig_thickness_Mh_all}a~and~\ref{fig_thickness_Mh_all}b).
In other words, at fixed $\mu(4h)$, $z_0(2h)/h$ for the thinnest model is
approximately equal to the value obtained from Eq.(\ref{eq_z-M_local}).

The data points lying on the lower envelope correspond to the models with
compact bulges or with massive halos. For such models, the bend was formed
in ``quiet'' regime or was suppressed altogether. In this case, the final
value of $\sigma_z / \sigma_R$ was also close to the saturation level of the
bending instability derived from the linear criterion. If, however, the
bending instability in the model grew in ``violent'' regime, then the final
value of $\sigma_z / \sigma_R$ could be considerably higher than that given
by the linear criterion and, hence, the final thickness of the model galaxy
could be much larger than that following from the linear criterion.

Thus, if a compact spherical component that can be represented by both a
bulge and a dark halo with a sharp central density peak (a cuspy halo) is
present in a real spiral galaxy in its central region, then the bending
instability in such a galaxy will grow in quiet regime. In this case, if the
galaxy is marginally stable against bending perturbations and perturbations
in the pane, then its parameters will obey Eq.~(\ref{eq_z-M_local}).

Consequently, we can use only relation~(\ref{eq_M-z}) to estimate the lower
limit for the dark halo mass from the relative thickness of galaxies. We
emphasize once again that this relation gives a weak constraint on the dark
halo mass, and only on the condition that the galaxy is in equilibrium, even
for thin galaxies.

\section{Discussion and conclusions}
It is well known that for a stellar disk to be stable against perturbations
in the plane, the value of $\sigma_R$ must be higher than a certain critical
value (Toomre 1964). On the other hand, the stability against the growth of
bending perturbations requires that the ratio $\sigma_z/\sigma_R$ be also
larger than a certain critical value. Consequently, the value of $\sigma_z$
is limited below for a stable stellar disk. The presence of a spherical
component, for example, a dark halo, has a stabilizing effect and reduces
the minimum radial velocity dispersion required for the stability against
the growth of perturbations in the plane. It thus follows that the disks
embedded in a massive dark halo can have lower values of $\sigma_z$ and, as
a result, can be thinner. These considerations underlie the conclusion about
the existence of a relationship between the relative thickness of marginally
stable stellar disks, $z_0 / h$, and the relative mass of the spherical
component, $\mu$, within a fixed radius (Zasov et al.1991, 2002). 

In this paper, we estimated the coefficient in this relationship and wrote
the corresponding theoretical relation~(\ref{eq_z-M_local}). The relationship
was confirmed by N-body simulations of disk galaxies. This theoretical
relation can be used to estimate the lower limit for the dark halo
mass from the observed thickness of real spiral galaxies~(\ref{eq_M-z}).
However, even for the thinnest galaxies, this relation yields a weak
constraint on the dark halo mass (the dark halo mass within four exponential
disk scale lengths must be higher than approximately one disk mass). We can
adduce several arguments that will relax this already weak constraint even
further.

First, we formally obtain an estimate of the lower limit on the total mass
of the galactic spherical component from Eq.~(\ref{eq_M-z}). The dark halo
gives only a partial contribution to this mass. Thus, the dark halo mass
estimate becomes even weaker. Second, as was already noted above, in
deriving Eq.~(\ref{eq_z-M_local}), we assumed that $\sigma_z/\sigma_R =
0.37$, which corresponds to the local linear criterion for bending
instability obtained by Polyachenko and Shukhman (1977). However, slightly
differing values of this ratio, within the range 0.29 to 0.37, were obtained
in different papers in which the bending instability criterion was derived
from a local linear analysis. We used the maximum theoretical value for the
dispersion ratio. If a smaller ratio is taken, then the coefficient in
Eq.~(\ref{eq_z-M_local}) will decrease and, as a result, the theoretical
estimate of the dark halo mass will also decrease.

Zasov et al. (1991) and Mikhailova et al. (2001) also derived the
relationship being discussed from $N$-body simulations of disk galaxies (in
general, the relationships given in these papers coincide). As was already
noted above, the model data points in these papers (in contrast to the
results of our simulations) lie well above relation~(\ref{eq_z-M_local})
and, as a result, the model relation at fixed stellar disk thickness predicts
a considerably higher mass of the spherical component than the theoretical
relation. What causes these differences? We see three causes.

The main cause of the differences is that the saturation level of the bending
instability in simulations can be much higher than the level given by the
linear criterion. This is probably the reason why the relationship being
discussed derived in the above papers lies above that found from the linear
stability criterion for bending perturbations. The numerical simulations by
Zasov et al. (1991) and Mikhailova et al. (2001) disregarded the fact that a
compact, but not necessarily massive, spherical component could effectively
suppress the bending instability. As a result, at fixed total mass of the
spherical component, the galaxies with a more concentrated spherical
component toward the center (e.g., if there is a compact bulge or a cuspy
dark halo in the galaxy) can be considerably thinner than the galaxies with
a looser spherical component. 

In addition, the overestimation of the model disk thickness is related to
the incorrect technique of determining the vertical scale height used in the
mentioned papers. It follows from our analysis that the characteristic disk
thickness $z_0$ in numerical simulations must be estimated via the median of
the absolute value of $z$ ($z_{1/2}$), not via the rms deviation of the $z$
coorrdinates of particles from the symmetry plane of the disk ($z_{\rm
rms}$). Otherwise, the disk thickness can be overestimated significantly, by
100\% or more. Thus, the mass of the spherical component determined from the
thickness of a marginally stable stellar disk will also be overestimated.
Note also that in the mentioned papers, the thickness averaged over the
entire disk, including its central regions, was taken when constructing the
experimental relationship. This disk-averaged thickness is definitely larger
than that calculated on the disk periphery, for example, at the distance $R
= 2h$ (for the radial profiles of the final model disk thickness, see
Mikhailova et al. (2001) and Sotnikova and Rodionov (2003, 2005)).
   
The ambiguity of the relationship between the thickness of a galaxy and the
mass of its spherical component with different density profiles in numerical
simulations complicates the estimation of the dark halo mass from the
relative stellar disk thickness $z_0/h$. However, to estimate the lower
limit for the dark halo mass from the relative thickness of real galaxies,
we can use theoretical relation~(\ref{eq_M-z}), which coincides with the
lower envelope of the numerical relationship
(Figs.~\ref{fig_thickness_Mh_all}a~and~\ref{fig_thickness_Mh_all}b). This
relation gives a weak constraint on the dark halo mass, $\mu(4h) \gsim 1$,
even for thin galaxies, which is consistent with the upper limits that
follow from considerations concerning the conditions for the existence of
spiral density waves in galaxies (Athanassoula et al. 1987) and from
arguments regarding the extent of tidal features in interacting galaxies
(Dubinski et al. 1999).


\vspace{0.5cm}
\noindent
{\large Acknowledgment}\\
\noindent
This work was supported by the Russian Foundation for Basic Research
(project no. 06-02-16459-a) and a grant from the President of Russia for
Support of Leading Scientific Schools (NSh-8542.2006.2).


\begin{center}
{\Large\bf References}
\end{center}

\medskip
\noi
S. Araki, Ph.D., Massachus. Inst. Tech., 1985.

\medskip
\noi
E. Athanassoula, A. Bosma, and S. Papaioannou, Astron. Astrophys {\bf 179},
23(1987).

\medskip
\noi
J. Barnes and P. Hut, Nature {\bf 324}, 446(1986).

\medskip
\noi
R. Bottema and J.P.E. Gerritsen, MNRAS {\bf 290}, 585(1997).

\medskip
\noi
J.N. Bahcall, Astrophys. J. {\bf 276}, 156(1984).

\medskip
\noi
F. Combes, R.H. Sanders, Astron. Astrophys. {\bf 96}, 164(1981).

\medskip
\noi
F. Combes, F. Debbasch, D. Friedli, and D. Pfenniger, Astron. Astrophys.
{\bf 233}, 82(1990). 

\medskip
\noi
J. Dubinski, J.Ch. Mihos, and L. Hernquist, Astrophys. J. {\bf 526}, 607(1999).

\medskip
\noi
L. Hernquist, Astrophys. J. Suppl. Ser. {\bf 86}, 389(1993).

\medskip
\noi
A.V. Khoperskov, A.V. Zasov, and N.V. Tyurina,
Astron. Zh. {\bf 78} 213(2001) [Astron. Rep. {\bf 45} 180(2001)].

\medskip
\noi
A.V. Khoperskov, A.V. Zasov, and N.V. Tyurina,
Astron. Zh. {\bf 80} 387(2003) [Astron. Rep. {\bf 47} 357(2003)].

\medskip
\noi
A.V. Khoperskov and N.V. Tyurina,
Astron. Zh. {\bf 80} 483(2003) [Astron. Rep. {\bf 47} 443(2003)].

\medskip
\noi 
M. Kregel, P.C. van der Kruit, and R. de Grijs, MNRAS {\bf 334}, 646(2002).

\medskip
\noi
R.M. Kulsrud, J.W-K. Mark, and A. Caruso, Astrophys. Sp. Sci. {\bf 14},
52(1971).  

\medskip
\noi
S.S. McGaugh and W.J.G. de Block, Astrophys. J. {\bf 499}, 41(1998).

\medskip
\noi
D. Merritt and J.A. Sellwood, Astrophys. J. {\bf 425}, 551(1994).
   
\medskip
\noi
E.A. Mikhailova, A.V. Khoperskov, and S.S. Sharpak,
Stellar Dynamics: from Classic to Modern
(Eds. L.P. Osipkov and I.I. Nikiforov),
St. Petrsburg: St. Petersburg State Univ. Press, 2001, p.~147.

\medskip
\noi
J. Navarro, C.S. Frenk, and S.D.M. White, Astrophys. J. {\bf 462}, 563(1996).

\medskip
\noi 
P.A. Patsis, E. Athanassoula, P. Grosbol, and Ch. Skokos, MNRAS {\bf 335},
1049(2002).

\medskip
\noi
D. Pfenniger and D. Friedli, Astron. Astrophys. {\bf 252}, 75(1991).

\medskip
\noi
V.L. Poliachenko and I.G. Shukhman,
Pis'ma Astron. Zh. {\bf 3}, 254(1977) [Sov. Astron. Lett. {\bf 3}, 134(1977)].
   
\medskip
\noi
S.A. Rodionov and N.Ya. Sotnikova, Astron. Zh. {\bf 82}, 527(2005) 
[Astron. Rep. {\bf 49}, 470(2005)].
   
\medskip
\noi
S.A. Rodionov and N.Ya. Sotnikova, Astron. Zh., 2006 (in press).

\medskip
\noi
J.A. Sellwood, Astrophys. J. {\bf 473}, 733(1996).

\medskip
\noi
J.A. Sellwood and D. Merritt, Astrophys. J. {\bf 425}, 530(1994).
   
\medskip
\noi
N.Ya. Sotnikova, V.P. Reshetnikov, Pis'ma Astron. Zh. {\bf 24} 97(1998) 
[Astron. Lett. {\bf 24} 73(1998)]
   
\medskip
\noi
N. Ya. Sotnikova and S. A. Rodionov, Pis'ma Astron. Zh. {\bf 29},
367(2003) [Astron. Lett. {\bf 29}, 321(2003)]
   
\medskip
\noi
N. Ya. Sotnikova and S. A. Rodionov, Pis'ma Astron. Zh. {\bf 31}, 17(2005)
[Astron. Lett. {\bf 31} 15(2005)]
   
\medskip
\noi
L. Spitzer, Astrophys. J. {\bf 95}, 325(1942).
   
\medskip
\noi
P.J. Teuben, ASP Conf. Ser. {\bf 77}, 398(1995).

\medskip
\noi
A. Toomre, Astrophys. J. {\bf 139}, 1217(1964).

\medskip
\noi
A. Toomre, Geophys. Fluid Dyn. {\bf 46}, 111(1966).

\medskip
\noi
T.S. van Albada, J.N. Bahcall, K. Begeman, and R. Sancisi, Astrophys. J.
{\bf 295}, 305(1985). 

\medskip
\noi
P.C. van der Kruit, Astron. Astrophys. {\bf 192}, 117(1988).

\medskip
\noi
A.V. Zasov, D.I. Makarov, and E.A. Mikhailova,
Pis'ma Astron. Zh. {\bf 17}, 884(1991) [Sov. Astron. Lett. {\bf 17},
374(1991)].

\medskip
\noi
A.V. Zasov, A.V. Khoperskov, and N.V. Tyurina,
Stellar Dynamics: from Classic to Modern
(Eds. L.P. Osipkov and I.I. Nikiforov),
St. Petrsburg: St. Petersburg State Univ. Press, 2001, p.~95.

\medskip
\noi
{A.V. Zasov, D.V. Bizyaev, D.I. Makarov, and N.V. Tyrina}
Pis'ma Astron. Zh. {\bf 28}, 599(2002) [Astron. Lett. {\bf 28}, 527(2002)].

\begin{flushright}
Translated by V. Astakhov
\end{flushright}

\end{document}